# Strengthening leverage of Astroinformatics in inter-disciplinary Science

M. Brescia*

*Department of Physics 'Ettore Pancini', University Federico II,*
*Napoli, Via Cintia, 21, I-80126, Italy*
*\*E-mail: massimo.brescia@unina.it*
*www.docenti.unina.it/massimo.brescia*

G. Angora

*INAF – Astronomical Observatory of Capodimonte,*
*Napoli, Salita Moiariello 16, I-80131, Italy*
*www.oacn.inaf.it*

Most domains of science are experiencing a paradigm shift due to the advent of a new generation of instruments and detectors which produce data and data streams at an unprecedented rate. The scientific exploitation of these data, namely Data Driven Discovery, requires interoperability, massive and optimal use of Artificial Intelligence methods in all steps of the data acquisition, processing and analysis, the access to large and distributed computing HPC facilities, the implementation and access to large simulations and interdisciplinary skills that usually are not provided by standard academic curricula. Furthermore, to cope with this data deluge, most communities have leveraged solutions and tools originally developed by large corporations for purposes other than scientific research and accepted compromises to adapt them to their specific needs. Through the presentation of several astrophysical use cases, we show how the Data Driven based solutions could represent the optimal playground to achieve the multi-disciplinary methodological approach.

## 1. Introduction

Let's start from an assumption: all sciences, and in particular Astronomy, are experimenting a drastic change of research perspective, in which the huge amount of data produced by modern telescopes and instruments are extremely heterogeneous and complex, causing the need to tackle the problem of processing them in an efficient way. That is why one of the main constraints is the parallel evolution of computing technology. We can compare a modern astrophysical survey to a Google search engine, in the sense that, like with Google, "research begins and ends with the survey itself!". Such a close parallel between Google and actual large surveys in astronomy refers to the modern survey properties: scale, quality, and richness of collected information. Scale, because we entered the era where we can observe and catalogue the entire sky. Quality, since source catalogues are as precise as the measurements taken with "pointed" observations. Richness, because those catalogues contain not only positions and fluxes, but also shapes, profiles, and temporal behaviour of the objects. This is what makes large surveys not just bigger

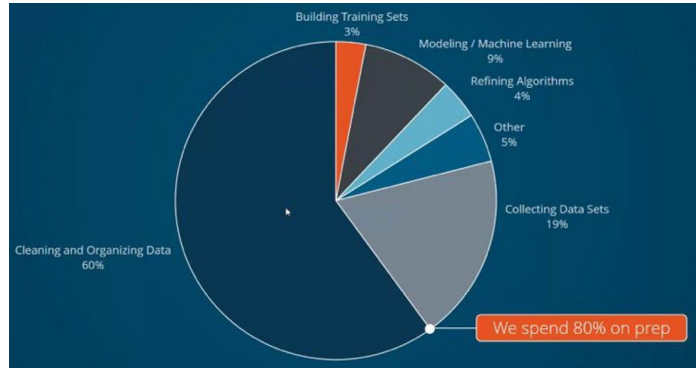

Fig. 1. Distribution of time data scientists spend across various experiment phases (source: forbes.com).

but better.

However, the real revolutionary aspect conditioning the science research evolution is that Astrophysics, like most all human sciences, is now facing the Big Data regime. Behind this, there is the awareness that the produced data volume and rate exceeds the capability to exploit and analyze them, thus causing data to become the driving engine of modern research. We like to remember that already 12 years ago, we were one of the few astrophysics research groups to publicly warn about the need to change strategy in the way of doing research, suggesting exchanging analysis models in large collaborations rather than data among the various research groups geographically spread across different countries [1,2]. At the time, we were almost ignored, especially within our national scientific community. Moreover, it is well known today that scientific research has changed its modus operandi. As shown in a recent research, we spend about 82% of our time dealing with data preparation, cleaning, understanding, cross-correlating, etc. (Fig. 1), well conscious that the quality of our scientific applications depends upon data accessibility, fusion, quality analysis, and durability when we want to apply Artificial Intelligence (AI) model to innovative problems on distributed heterogenous archives.

Formally, a Big Data domain is defined as an environment characterized by the so-called 8-V rule: Volume, Value, Veracity, Visualization, Variety, Velocity, Validity, and Variability [3]. Given its intrinsic nature, it is evident that Astrophysics perfectly matches such a definition.

A Big Data environment also has several critical properties fundamentally rooted in the nature of data itself. Within this context, a datum can be defined as a piece of information, a reference point of a scale or operation, an assumed or known entity, a factual basis for knowledge inference, or an element for reasoning. These definitions can be collectively encapsulated in a single word: a datum is a "solution". That is, each datum is a solution to its domain problems; thus, in a Big Data regime, we have an order of $\sim 10^{20}$ solutions, each one in a multi$-$D parameter space. And if we think that in the observable Universe, the order of galaxies should be $\sim 10^{12}$, the conclusion is around the corner. Therefore, if we can collect enough data in a hyper-dimensional parameter space, we will obtain a statistically consistent model for that environment. This means that all the questions concerning the research in that field are already present in the collected volume of data. The problem is just to find them! The immediate consequence is that a Big Data domain requires enough computing and storage power to perform an efficient data cross-correlation and a massive exploitation of self-adaptive analysis mechanisms, able to automate the data exploration process, or, in other words, AI learning paradigms.

## 2. ICT infrastructure is needed

Over the last 40 years, research in Astrophysics has undergone a gradual evolution comparable to that of particle physics and many other fields, starting from medium-small scale projects (e.g. SDSS), characterized by a single set of instrumentation, considered as cutting-edge technology at the time, by relatively low budgets with a medium-high risk rate, and with the possibility of easily introducing new AI methodologies, but soon limited by the growing need for computational power. Within two decades, large-scale international collaborations (such as ESA Euclid, Rubin LSST, JWST, ELT, and SKA, to quote just a few) centered on big science experiments (several hundred participants), with large budgets and a lower risk rate, but based on hardware and software technologies frozen at an early stage and computational infrastructures on a national scale, managed in the form of large data centers that were not particularly flexible or easily innovative. These are all projects that cover the entire lifetime of a researcher, where AI, Machine Learning (ML), and Deep Learning (DL) methodologies act mainly as support at all levels and where large volumes of data require a long-term maintenance cycle, with a large fraction of the costs to be invested in data preservation and curation.

Therefore, right from the start, computation and storage problems emerge around the big science initiatives, where hardware plays a central role with almost 90% of the total cost. Fortunately, technological evolution and new energy policies make hardware a commodity that is easily available, providing an underlying infrastructure capable of focusing attention on investments in data management and efficient software production. Naturally, a relevant aspect is that of sustainable consumption. It is well known that maintaining the data and the hardware/software infrastructures produces yearly more $CO_2$ than the entire airline industry. Just think that training one medium-sized AI model consumes about 300 kilos of $CO_2$, while a Large Language Model reaches almost 10 tons of $CO_2$ [4]. This implies that a Tier-II data center consumes more energy than that needed to power an urban core of 50 thousand apartments, although it is still cheaper than 10 smaller data centers (the so-called economy of scale) [5]. This is why the new mantra[a] of the European

---

[a]https://digital-strategy.ec.europa.eu/en/policies/green-cloud

Union is "minimize the proliferation of data computing centers and create sustainable models of computing infrastructures for the management of large volumes of data".

In such a scenario, once again, Astroinformatics and, more generally, X-informatics can play a fundamental role, offering a level of knowledge and flexibility capable of combining the maximum and efficient exploitation of computational resources with the ability to manage and synthesize the analysis of massive amounts of data, maximizing scientific production. The guiding element of the new computing data centers must be the multi-disciplinary management by data scientists, specialized in the different scientific domains (bio, astro, geo, chemio, etc.) to which the hosted data refer. The reasons are many. First, it is unnecessary to rediscover the wheel every time. In data-intensive projects, there are often many repeated patterns with respect to different scientific disciplines, so it is advisable to turn these patterns into processes that can be replicated. Furthermore, large data sets are difficult to aggregate, so for multi-disciplinary data centers, it is preferable to exploit the co-location mechanism, which would be impossible for exclusively astronomical or generally mono-thematic data centers, given that the ontologies and integrated data models are complex combinators. The co-location principle allows the optimization of the aforementioned mechanism of a sustainable model for massive data storage over many decades and computing infrastructures. Of course, by respecting the constraints of the need to find the right balance between HPC and HTC resources, to guarantee active support for the creation of new AI-based models and therefore to recognize the costs related to the design and production of AI solutions as a significant portion of the entire big science project.

That is why we profited from the recent "Next Generation EU" funding program of the European community, to provide a state-of-the-art data computing center to make our multi-disciplinary community able to exploit the incoming tsunami of data, in particular for Astrophysics collected by a variety of survey projects (to quote just a few ELT, SKA, Rubin-LSST, VST). This new infrastructure, named AD-HOC (Astrophysical Data HPC Operation Center) and part of the Italian project STILES[b] (Strengthening the Italian leadership in ELT and SKA), will be operative from the end of 2024, offering a TIER-II data center reaching two petaflops of computing capability, equipped by both HPC and HTC server clusters, together with an overall data storage capacity of about eighteen Petabytes, between solid-state disks and long-term archiving tape libraries [6]. Among the various services foreseen, it is the candidate to be the official archiving and processing system for specific survey projects, like VST.

---

[b]https://pnrr.inaf.it/progetto-stiles/

## 3. AI as the engine of data-driven Science

Nowadays, AI is recognized as an unavoidable approach in all human sciences. However, mainly in the academic and basic research world, in many cases, the enormous vastness and variety of software resources and solutions based on AI and machine learning available on the web, combined with the widespread indiscriminate and often only partially conscious use of such resources, have begun to highlight not only the merits but also the limits and critical issues of such methodologies. The strong pressure exerted on experts in the various scientific domains by the ever-increasing role of data science experts involved in large scientific collaborations has certainly contributed to amplify this effect. This, rather than encouraging synergy between heterogeneous scientists, has caused the triggering of a paroxysmal race among colleagues and members of the same project to obtain barely publishable results in a short time, forcing the use of AI techniques and exponentially increasing their complexity to the point of losing control, making them almost incomprehensible and difficult to interpret and to become trustable. The tendency is, therefore, that the primary objective of doing the best possible science, the central concept on which the AI approach is based, i.e. data-driven science, as well as the spirit of cooperation between domain and AI experts are failing. In practice, the key objective of improving knowledge in the scientific field does not represent, in many cases, the most important target of research, worrying only about pursuing subjective goals linked to personal prestige or acquiring research funds. This distortion of the scientific perspective is particularly and dangerously present in Italy, where national support for basic research is chronically insufficient. The final result is that AI has become a land of conquest with little awareness and without rules. But all that glitters is not gold. The AI approach presents various critical issues that should be considered before deciding to use it in any scientific field. In particular, the concept of data-driven science is often forgotten or underestimated. The in-depth critical analysis and knowledge of the data underlying any experiment should be the key element to focus attention most of the time, especially before deciding to employ any machine learning-based model.

For example, in Astrophysics, the massive presence of AI is demonstrated by the exponential increase in scientific production (Fig. 2). Consequently, there is an increasing need for an efficient set of rules to build and use this methodology efficiently and to report their results as well [7]. As known, in the last decade, it became a common practice to engage data challenges within large survey programs to find the best solutions to analyze data [8–13]. But most suffer from inhomogeneous rules and objective metrics to validate results [14].

In principle, four main criteria should be followed by any AI approach in human science: (i) Reproducibility, in the sense that the results should be invariant w.r.t. data training/prediction choice and code/data made publicly accessible; (ii) Interpretability, because methods should be explainable and transparent in terms of both pipeline and data flow process; (iii) Accuracy of methods, since they should provide accurate results, i.e. unbiased, uniform and rigorous in terms of their validation; (iv) Flexibility, for which AI solutions and approaches should demonstrate general usefulness and non-ambiguous advantages w.r.t. traditional approaches.

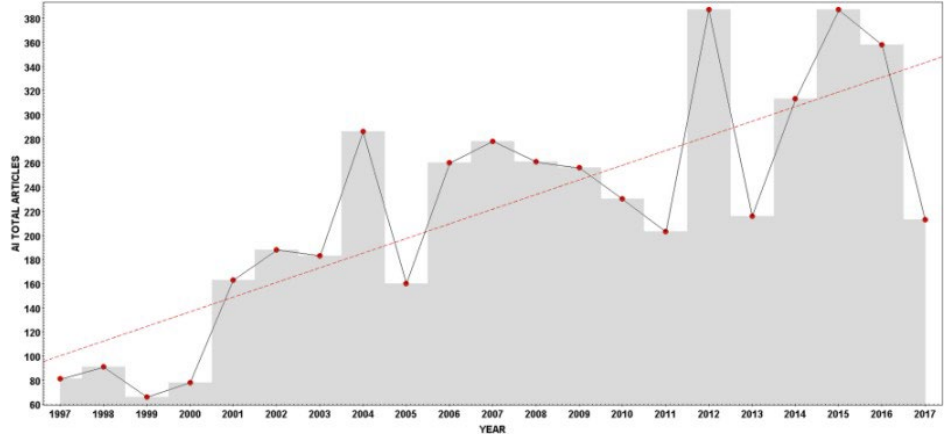

Fig. 2. Amount of scientific articles focusing on Astroinformatics, published per year on specialized refereed journals, from January 1997 to July 2017[15].

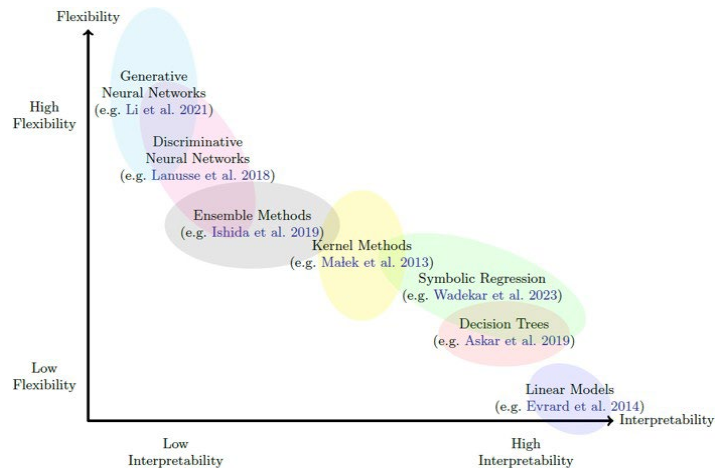

Fig. 3. Trade-off between interpretability and flexibility for AI based methods[7].

Unfortunately, some of these criteria conflict with each other. For example, for their intrinsic design aspects, the degree of flexibility of Machine and Deep Learning models is inversely proportional to their level of interpretability, as shown in Fig. 3.

In practice, simpler methods, like linear models, cannot be adaptable to different tasks despite their intrinsic transparent mechanisms. On the opposite, the recent generative models showed a very high capability to solve different complex problems but with poor levels of interpretability. A trade-off should be found, trying to follow what the data themselves suggest and, as much as possible, simpler in the solving approach, increasing the complexity of the model only if needed. Anyway, as easy to understand, this trade-off is extremely problem-dependent.

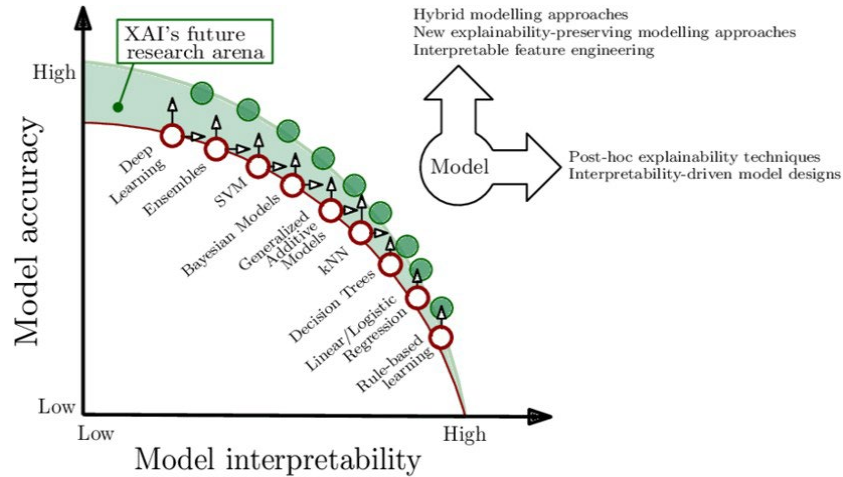

Fig. 4.  Trade-off between accuracy and interpretability for AI based models [16].

Moreover, as we are assisting worldwide in many contexts, the debate on the ethics of generative AI is controversial and often characterized by poor objectivity and misunderstandings. The main reason is that it appears extremely accurate in solving very hard problems but at the price of a very low capability for a human being to understand how it works and how to keep it under control. The EU recently published a GDPR law, known as the AI Act[c], focusing on these aspects and trying to regulate its evolution and diffusion. But, as shown in Fig. 4, also, in this case, there is an inverse proportionality between accuracy and explainability.

To be trusted and accepted by a scientific community, the internal behaviour of a model should be clearly understood. Recently a new branch of research in AI arisen exactly to introduce well-posed mechanisms and rules able to increase the explainability of machine learning models, the XAI or eXplainable AI. Several methods have been recently proposed [16], and in particular, one of them, SHAP, includes an efficient and reliable way to make the feature selection problem fully explainable and therefore trustable [17].

---

[c]https://www.euaiact.com/

### 3.1. *Parameter space explainable optimization*

As known, in a data-driven context, one of the most important aspects to consider in a scientific problem is a deep exploration and analysis of the given parameter space on which the available data are mapped. By definition, ML is based on self-adaptive techniques. A priori, real-world data are intrinsically carriers of embedded information hidden by noise. In almost all cases, the signal-to-noise (S/N) ratio is low, and the amount of data prevents human exploration. We don't know which feature of a pattern is carrying useful information and how and where the correlation of features gives the best solution knowledge. In Astrophysics, this particularly means that before approaching any AI solution, a massive investigation should be dedicated to understanding the degree of correlation among all data features and the level of information contribution carried by each single dimension. But almost always, this means efficiently analyzing a high dimensional parameter space characterized by many features, thus subject to the well-known problem of the curse of dimensionality [18]. Principal Component Analysis (PCA) [19], the most widely used method, is very simple but often inefficient. In particular, after having extracted the principal trends of data distribution, all the rest is considered as noise contribution, thus causing a lack of residual information that, in several complex domains, could be crucial to fully exploit the benefits of the feature selection to optimize the solution approach and to infer new knowledge on the underlined physical phenomena. Namely, rejected components may hide a weak but still relevant contribution to better understanding data.

There is an alternative method to PCA. A few years ago, we proposed a hybrid method named PHILab (Parameter Handling Investigation Laboratory), mixing two existing models and joining two simple concepts, for instance, shadow features and Lasso regularization, successfully validated on particularly complex and multi-dimensional astrophysical use cases [20–22]. A shadow feature, proposed by the Boruta method [23], is a copy of a real one, present in the original parameter space, built by randomly shuffling its values. This corresponds to introducing quotes of random noise into the given dataset. Lasso regularization [24], based on the L1-norm additive term added to any loss function, is able to shrink the informative correlation between features, making them candidates to be rejected from the parameter space, as being considered redundant because they already sufficiently represented by other features. Therefore, Boruta can classify all real features as relevant, rejected, or undetermined. The undetermined features are submitted to the Lasso analysis to make their final assignment. The real advantage of this method is its demonstrated reliability in finding the full set of features carrying an informative contribution, however weak, to the problem solution, thus improving the completeness of the feature selection process and potentially engaging serendipity.

Although the cited approaches can improve the quality of the feature selection, these methods are not fully explainable due to two main limitations: the intrinsic variability of the feature importance calculation obtained by the tree-based bag-ging/boosting models used as engines by the mentioned methods, as well as their low explainability of the selected parameter space, due to their high dose of randomness and statistical complexity. An XAI model, like SHAP, could solve both problems.

SHAP is a framework inspired by the game theory [17], where the easy concept is that a parameter space feature may or may not be involved in the game. When involved, its contribution to the model performance is evaluated by comparing it with all the other features. However, since the order of participation of a set of features may be important, all possible combinations of features are compared to avoid such a bias. This full comparison among features and all the dataset samples provides the possibility to estimate at the same time three kinds of informative contributions for each feature: the cumulative effect of a feature with respect to the entire dataset, the main effect of a single feature for a given data sample, and the interaction effect for all pairs of features for a given data sample. Therefore, SHAP provides an easy and interpretable evaluation of the information contribution carried by each feature of the parameter space on its own and in correlation with all the others. At the same time, it also improves the feature selection reliability and the explainability of the internal estimation mechanisms (see Fig. 4).

Recent applications compare the feature selection methods mentioned above on astrophysical use cases characterized by large volumes of data represented through multi-dimensional parameter spaces. For example, the classification of the evolutionary stage of Young Stellar Objects (YSOs) through the characterization of the Hi-GAL (Herschel InfraRed Galactic Plane Survey) clumps parameter space, trying to infer new insights on the connection between the cold material reservoir present in clumps, traced by FIR/sub-mm emission, and the already formed YSOs, precursor of stars [25–27]

When the classification performances are compared between the full original parameter space and that extracted by these methods (Fig. 5), a series of important outcomes of the feature selection can be outlined: (i) it does not lose information and by drastically reducing the dimensions of the problem it speeds up the process; (ii) in some cases it improves the classification quality, mostly in terms of the best trade-off between purity and completeness; (iii) the SHAP mechanism is intrinsically fully explainable; (iv) the importance measured for the features may provide potentially interesting aspects to help the post-processing analysis of results in terms of star-forming evolution.

Another example is the comparative study of different cosmological models on Pantheon+SH0ES data[d], with the primary aim of evaluating the robustness of the $\Lambda$CDM model with respect to other dark energy models and investigating whether there are deviations that could indicate new cosmological insights. The data-driven approach through the various feature selection techniques (in particular Boruta and SHAP) allows us to evaluate the informative contribution of the cosmological and statistical parameters derived from the light curves to derive the supernova distance modulus. The resulting parameter space extracted from the feature selection models reveals potential improvements in alternative models to the $\Lambda$CDM, relevant in perspective for new observational campaigns, as in the case of the recent DESI survey [28,29]. These use cases demonstrate the effectiveness and the centrality of the data-driven approach, realized through feature selection models, to improve the cognitive inference on the related physical phenomena, combined with the possibility of ensuring the full explainability and transparency of the method.

---

[d]https://zenodo.org/records/4015325

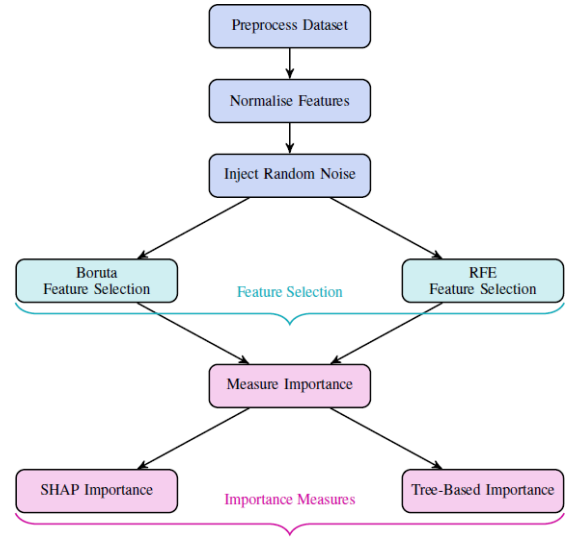

Fig. 5. An example of typical data-driven processing flow to perform the feature selection on a multi-dimensional parameter space [27].

## 4. Astroinformatics as a virtuous inter-disciplinary synergy template

Besides interpretability, two other important properties should characterize the use of AI within scientific research: accuracy and flexibility. Several examples could be done to verify these properties. The first is the strong parallelism between multi-spectral analysis in astrophysics and biomedicine. By thinking about the domain of interest of Astrophysics, we know that this is a science mostly oriented to describing and explaining the evolution of the Universe by recognizing and characterizing all celestial sources. We daily collect and use multi-resolution, multi-epoch, and multi-wavelength images and spectra, measuring several physical quantities from the extracted source catalogues, and we already demonstrated that this is done in a Big Data environment. Perfectly matching this modus operandi is the high-resolution spatial transcriptomics, a branch of biomedicine, where the same kinds of massive data, but referred to biological tissue cells (here we can consider cells as the galaxies of the biological universe), are analyzed to describe and explain the evolution of human being organisms and to recognize and characterize all involved and interacting cells. The well-established technique of RNA sequencing allows us to obtain information on the state of cells, tissues, and entire organisms at various time points in the form of gene expression profiles. Such in-depth knowledge is particularly useful in identifying pathologies and their evolution based on the type of treatment adopted and on particular environmental conditions. Therefore, strong similarities emerge between multispectral analysis in astrophysics and the emerging multiplexing platforms in biomedical research. The unstoppable sequential identification of tissue-based biomarkers makes this scientific field a big data environment. Therefore, it seems clear that image analysis techniques originally designed and tuned for problems in astrophysics can be applied to the biomedical field to improve tissue imaging methods, facilitating their efficiency and reliability of identification, prediction, and classification [30].

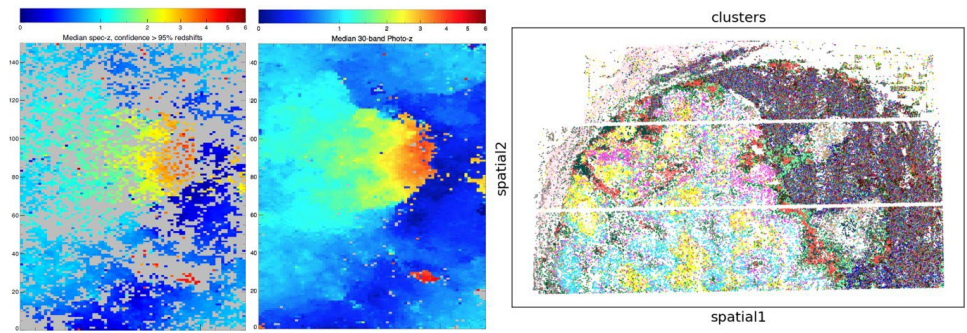

Fig. 6. Examples of clustering analysis. On the left, two sub-panels show a SOM output of a distribution of thousands of galaxies grouped in terms of median spectroscopic redshift (left sub-panel) and of median 30-band photometric redshift (right sub-panel) [31]. On the right, a nonlinear PCA output, based on an autoencoder, shows a distribution of different clusters of cells in a mouse lymph node tissue frame [32].

In both scientific domains, the data are mostly based on multi-resolution, multi-epoch, multi-band images. This strong parallelism automatically involves the same approach of AI methods. It can, therefore, give rise to synergies and collaborations between different and apparently distant scientific disciplines, whose main goal is to exploit the experience acquired in the parameter space analysis and classification and whose baseline is the exploitation of accuracy and flexibility properties of AI paradigms. For example, trying to be data-driven in the recognition of hidden structures within the two different parameter spaces, the clustering approach reveals strong analogies.

For instance, in Fig. 6, on the left, we can see the output of a Self Organizing Map (SOM) model, consisting of a distribution of dozens of thousands of galax-ies grouped in terms of median spectroscopic and 30-band photometric redshift, providing an effective way to compare the spectroscopic and photometric coverage of the parameter space [31]. On the right, a modified version of the clustering step of the Open-ST pipeline [32], showing the distribution of different clusters of cells in a mouse lymph node tissue frame obtained after the application of a Nonlinear PCA (an autoencoder), propaedeutic to discover the potential presence of diseases along the tissue evolution analysis. In practice, different science domains with the same methodology and comparable reliability of results, thus probing the general accuracy and flexibility of the AI approach.

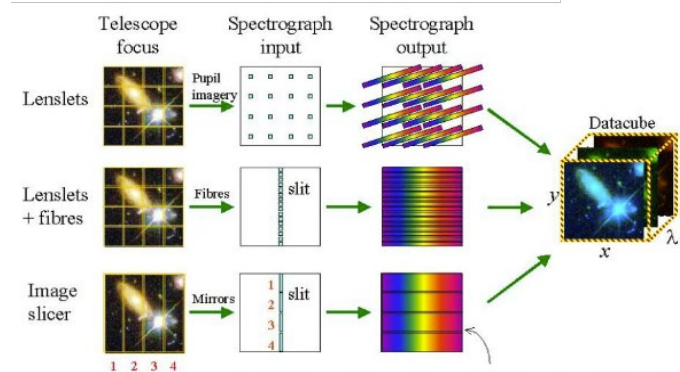

Fig. 7.  Operating principles of Integral Field Spectroscopy[34].

Another example is the strong parallelism between Astrophysics and Geophysics. We know that in Astrophysics, the integral field spectroscopy is particularly suitable to characterize the morphology of celestial sources (Fig. 7) by performing an accurate estimation of position and energy emitted by single photons[33].

A similar kind of investigation can be performed in Geophysics by remote sensing techniques (Fig. 8) to analyze and characterize the chemical and geological composition of the Earth and other planets[35].

In both cases, Hyperspectral Imaging (HSI) can provide a continuum spectrum of light, thus achieving an accurate estimation of the physical nature of the observed objects, because it records a continuum spectrum of light for each pixel and provides an invaluable source of information regarding the physical nature of the different object compositions, leading to the potential of a more accurate classification. The advent of Deep Learning has revolutionized the field of HSI, particularly from the introduction of Visual Transformer (ViTs) models [36]. A ViT may capture features at multiple scales and demonstrate strong generalization across different types of multi-band images. We recently proposed a customized version of these models, which extends the self-attention mechanism to 3D image patches, AMBER, an advanced SegFormer for Multi-Band Image Segmentation.

As shown in Fig. 9, it consists of two main modules: a Transformer encoder,

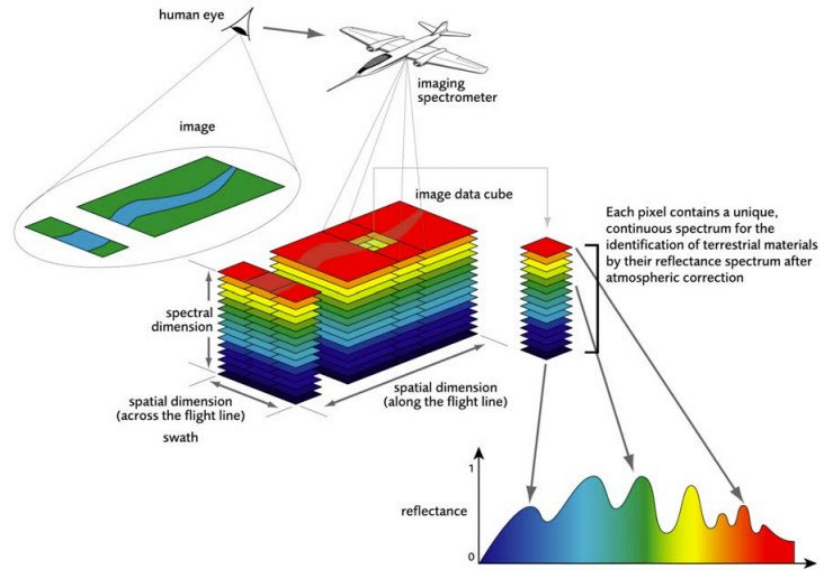

Fig. 8.   Operating principles of geophysical remote sensing [35].

which generates both high and low-resolution features, and a lightweight All-Multi-Layer-Perceptron decoder (All-MLP) used to fuse multi-level features together. In particular, the final layer of the decoder reduces the dimensionality to generate a bi-dimensional semantic segmentation mask from a 3D input image. Unlike the traditional models, AMBER is tailored to multi-band images due to its 3D convolution. Moreover, the proposed model preserves the input spatial dimensions to maximize the accuracy at the price of a relative increase in the number of trainable parameters.

Preliminary experiments conducted on the Indian Pines, Pavia University, and PRISMA datasets show that AMBER outperforms traditional CNN-based methods in terms of Overall Accuracy, Kappa coefficient, and Average Accuracy on the first two datasets, achieving state-of-the-art performance on the PRISMA dataset [37]. In Fig.10, there is an example of an application on remote sensing problems, where randomly extracted patches of the training images were used to test the capability of the SegFormer to recognize different kinds of regions, resulting better than other models proposed in the literature.

## 5. The critical optimization of the AI model

Another important aspect concerns choosing and optimizing the proper model for any particular scientific use case. Usually, an approach to problem-solving with AI should tackle two main aspects: the optimization trade-off between model and data domain parameter spaces and the computing efficiency in the problem-solving heuristics. In fact, the optimization of the parameter space is only one aspect to take into account. Other equally important factors are the right choice of the best model, the optimization trade-off between its hyperparameters and the data feature space, and the heuristic search for the optimal AI architecture and depth. In all cases, it is always extremely data-dependent, which is why AI methods should be approached in a data-driven way. One efficient way to accelerate the optimization of the right model is through a special learning paradigm called meta-learning. This type of ML focuses more on the learning process than on the final result of any experiment (Fig. 11). In traditional machine learning, a model is trained on a specific dataset to perform a specific task. In meta-learning, a model is trained on a set of tasks to learn quickly how to adapt to new tasks in the future.

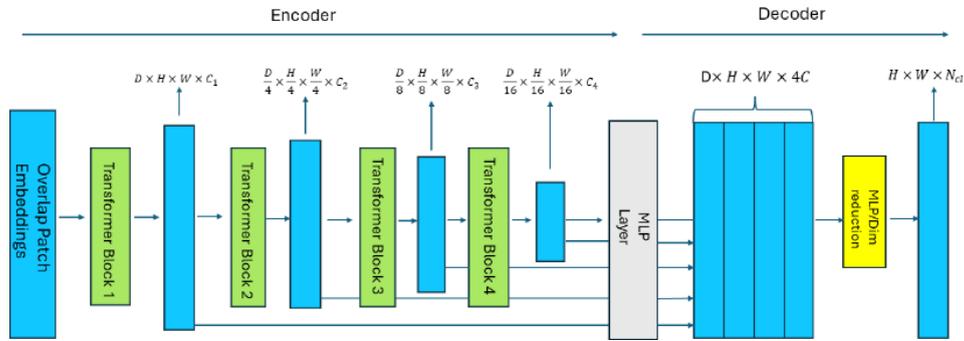

Fig. 9. The architecture of the AMBER ViT consists of two main modules: A hierarchical transformer encoder to extract spatial/spectral features and an All-MLP decoder to fuse these multi-level features, reducing the spectral dimension and predicting the semantic segmentation mask. D, H, W, and $C_i$ represent the image deepness (spectral dimension), image height, width, and the features of the three-dimensional convolution, respectively [37].

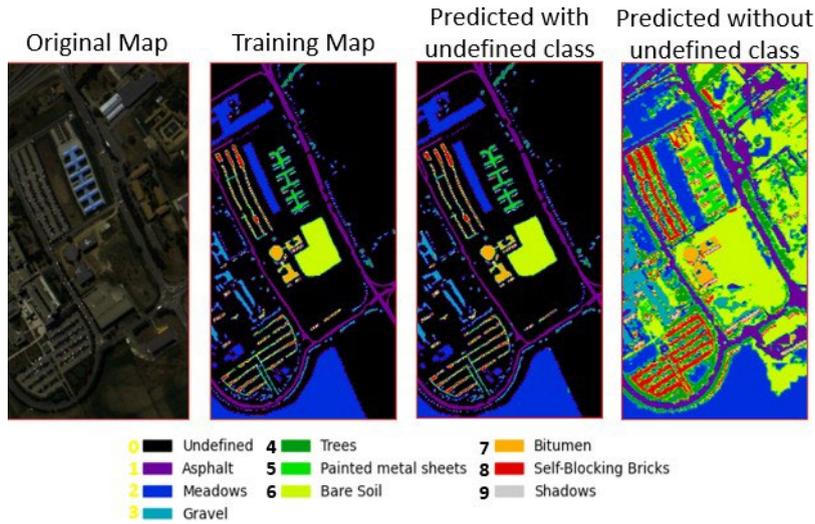

Fig. 10. AMBER prediction on Pavia University dataset. From left to right: False-color map, Ground-truth map, AMBER prediction with the undefined mask, and the AMBER prediction without the undefined mask [37].

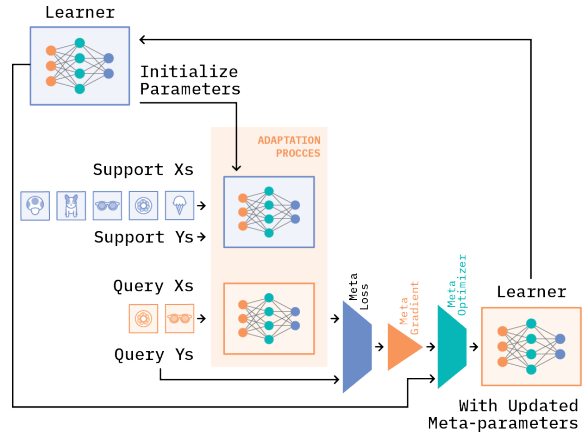

Fig. 11.  Meta learning process (source: https://meta-learning.fastforwardlabs.com).

In analogy with how humans exploit the acquired experience to learn how to solve novel problems, a meta-learning process is based on a double training loop, where a model is trained on several (usually few) data for a series of tasks, and then it leverages its experience to improve its learning on novel tasks. For example, a model could be trained to reproduce input image data, and afterward, it can learn how to classify objects inside them. At first glance, this concept may appear already known to the eyes of a machine and deep learning expert since it seems very similar to the transfer learning paradigm [38]. However, it should be kept in mind that in the case of transfer learning, a pre-trained model requires a huge amount of data and a considerable computational capacity. Furthermore, such technique may suffer from some limitations in problems on specific scientific domains, such as looking for rare or faint astrophysical sources in very limited S/N conditions, where an equally large amount of labeled data should be needed to obtain good reliability [39].

The meta-learning paradigm is particularly interesting for its ability to learn from a few data, adapt to new tasks, and its propensity to a high degree of generalization. These aspects make it particularly useful in specific scientific applications for which the acquisition of large amounts of available labeled data is expensive or difficult. In fact, a meta-learning process is essentially based on a first training level, followed by an adaptation level. In the first step, a model learns a preliminary set of general parameters on a wide set of tasks and then focuses on reusing the acquired experience to learn a specific set of parameters on a new task.

In particular, here we would like to focus on a promising optimization technique of meta-learning called Model Agnostic Meta Learning or MAML [40,41]. It is based on the idea of training a model to be easily adapted to new tasks with just a few optimization steps. This occurs in a model-agnostic way through an update of the initial parameters using a reduced number of training iterations with respect to the new task in the meta-learning phase. The intuition is that learning from an already sufficiently good initialization to improve the model preliminarily will allow us to obtain better efficiency in the generalization phase. The MAML approach is extremely versatile and usable for various heterogeneous problems. It is also applicable to almost all main machine learning paradigms, and it is able to adapt to any type of deep learning model.

For example, the MAML paradigm was recently applied to a particularly complex problem in astrophysics, for instance, the radio interferometry deconvolution problem, or the reconstruction of the true sky from the observed one on massive data cubes (a sample of 1000 cubes, each one with a size of about 100 GB, 10 M spatial features and 10 K frequency channels) referred to the ALMA project data [42–44]. As a usual practice in AI, the best model should be found by performing a computing-intensive campaign to find the optimal setup of the model. This process depends on the number of internal hyperparameters directly dependent on the model's complexity. We know that a typical DL model comprises thousands of internal parameters, and there could be dozens of models to explore. The proposed framework DeepFocus is a MAML approach based on the Maximum Likelihood Estimation (MLE) technique applied to the model space, which finds the optimal combination of feature extraction and optimizer parameters in a double training loop, massively exploiting the GPU accelerators to speed up the search. This exactly follows the model agnostic meta-learning paradigm (Fig. 12).

An input data cube is used in an inner training loop to identify the better feature extraction model architecture, allowing for a guess of the sky model behind the dirty input data. Then, the encoder part, which stores the extracted features in the latent space, is reused by attaching a classifier to extract the interesting sources from the proposed sky model image, which are compared with the target ones using a supervised approach. Through the MLE-based loss function, the internal weights of the model are updated, and a new double training loop is executed. Therefore, DeepFocus explores the model space at each double loop using the acquired experience on the sky model image reproduction learning to optimize the classifier learning. It searches for the best DL model by maximizing the likelihood of efficiently exploring the model architecture space on top of an HPC computing infrastructure. It reveals a high capability to find the best trade-off between accuracy and flexibility [43]. DeepFocus was compared with several methods, among which the traditional approach, based on the well-known TCLEAN pipeline [45], which approximates the deconvolution problem, using a massive number of physical priors, including several models of the expected noise sources, continuum subtraction, etc. We tried to use

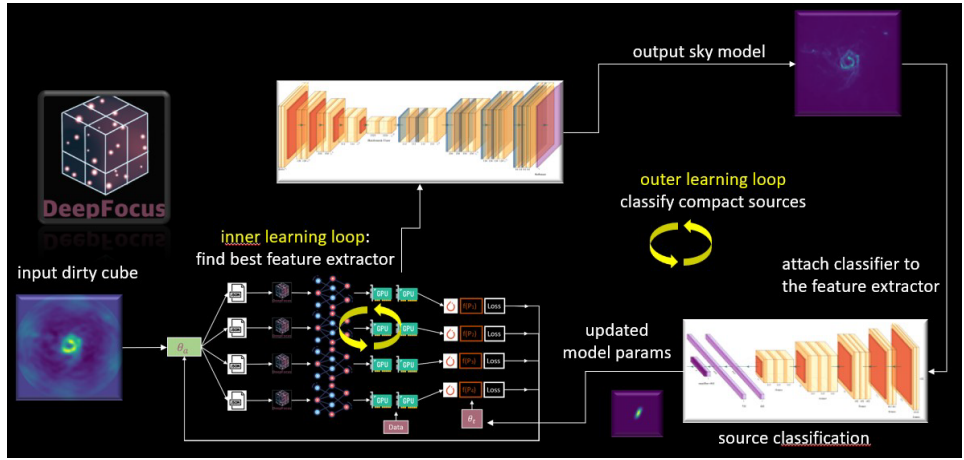

Fig. 12.   DeepFocus MAML functional process flow.

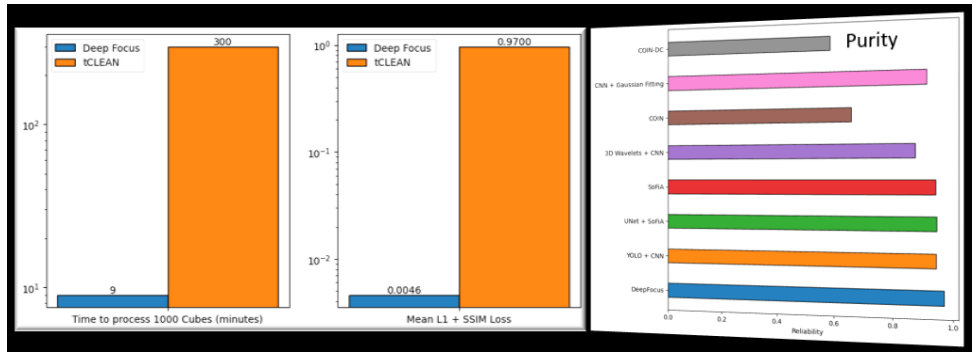

Fig. 13.   DeepFocus accuracy and speed up performances.

our DeepFocus to both speed up the process and gain accuracy.

As shown in Fig. 13, after a test campaign done on 1000 datacubes, DeepFocus was able to drastically reduce the processing time, being able to digest 100 TB of data in 9 minutes, rather than the 5 hours required by the TCLEAN pipeline. It also improved the classification precision compared to other solutions proposed in the literature, some of which are also based on ML/DL approaches.

## 6.  Concluding remarks

The fundamental concept arising from all we have discussed is the unfeasibility of the classical approach to scientific investigation based on the so-called hypothesis-driven paradigm. The Big Data scenario imposes a new methodology, fully centered on the rule to be driven by data themselves at all levels of the scientific approach. All sciences can now be considered cyber-science (a.k.a. e-science), which requires new scientific methodologies. The challenges we are tackling, invariant to any specific science field, mainly concern managing and exploring large, complex, distributed data to engage serendipity and a virtuous synergy between domain-specific and data-driven science. In such a scenario, AI's central and unavoidable role must be regulated globally but positively by thinking that, as for all human discoveries, its impact on humankind's evolution depends only on us.

**Acknowledgments**

We would like to underline that all the contents of this work would not be possible without the huge and constant contribution of a fantastic pool of collaborators (better friends) coming from different Institutions and a variety of scientific and technological fields: Stefano Cavuoti, Maurizio D'Addona, Mariarca D'Aniello, Michele Delli Veneri, Carlo Donadio, Andrea Dosi, Adriano Ettari, Antonio Ferragamo, Giuseppe Longo, Ylenia Maruccia, Amata Mercurio, Fabio Ragosta, Giuseppe Riccio, Alvi Rownok, Guido Russo, Maria Zampella. They are the core team, with the standard meaning of the core term in English, especially with the Neapolitan meaning 'o Core team (Hearth Team).

August 30, 2024 9:23 ws-procs961x669 WSPC Proceedings - 9.61in x 6.69in output page 2020

27. Y. Maruccia, et al., in prep.
28. S. Vilardi, S. Capozziello and M. Brescia, Discriminating among cosmological models by data-driven methods, *arXiv e-prints, Submitted to A&A*, p. arXiv:2408.01563 (August 2024).
29. DESI Collaboration, A. G. Adame and al., DESI 2024 VI: Cosmological Constraints from the Measurements of Baryon Acoustic Oscillations, *arXiv e-prints*, p. arXiv:2404.03002 (April 2024).
30. A. S. Szalay and J. M. Taube, Data-Rich Spatial Profiling of Cancer Tissue: Astronomy Informs Pathology, *Clinical Cancer Research (2022* **28**, 3417 (August 2022).
31. D. Masters, P. Capak, D. Stern, O. Ilbert, M. Salvato and al., Mapping the Galaxy Color-Redshift Relation: Optimal Photometric Redshift Calibration Strategies for Cosmology Surveys, *ApJ* **813**, p. 53 (November 2015).
32. M. Schott, D. Leon-Periñán, E. Splendiani, L. Strenger, J. R. Licha and al., Open-st: High-resolution spatial transcriptomics in 3d, *Cell* **187**, 3953 (2024).
33. R. Bacon, J. Brinchmann, S. Conseil, M. Maseda, T. Nanayakkara and al., The MUSE Hubble Ultra Deep Field surveys: Data release II, *A&A* **670**, p. A4 (February 2023).
34. J. R. Allington-Smith, Integral Field Spectroscopy for Panoramic Telescopes, in *Revista Mexicana de Astronomia y Astrofisica Conference Series*, ed. S. Kurtz (Revista Mexicana de Astronomia y Astrofisica Conference Series, June 2007).
35. E. Knaeps, S. Sterckx, M. Bollen, K. Trouw and R. Houthuys, Operational remote sensing mapping of estuarine suspended sediment concentrations (ormes) (01 2006).
36. A. Dosovitskiy, L. Beyer, A. Kolesnikov, D. Weissenborn, X. Zhai and al., An Image is Worth 16x16 Words: Transformers for Image Recognition at Scale, *arXiv e-prints*, p. arXiv:2010.11929 (October 2020).
37. A. Dosi, M. Brescia, S. Cavuoti, M. D'Aniello, M. Delli Veneri, C. Donadio, A. Ettari, G. Longo, A. Rownok, L. Sannino and M. Zampella, AMBER - Advanced SegFormer for Multi-Band Image Segmentation: an application to Hyperspectral Imaging, *Submitted to Neural Computing and Applications*.
38. F. Zhuang, Z. Qi, K. Duan, D. Xi, Y. Zhu, H. Zhu, H. Xiong and Q. He, A comprehensive survey on transfer learning (2020).
39. S. Cavuoti, D. De Cicco, L. Doorenbos, M. Brescia, O. Torbaniuk, G. Longo and M. Paolillo, Identification of problematic epochs in astronomical time series through transfer learning, *A&A* **687**, p. A246 (July 2024).
40. C. Finn, P. Abbeel and S. Levine, Model-Agnostic Meta-Learning for Fast Adaptation of Deep Networks, *arXiv e-prints*, p. arXiv:1703.03400 (March 2017).
41. Y. Tian, X. Zhao and W. Huang, Meta-learning approaches for learning-to-learn in deep learning: A survey, *Neurocomputing* **494**, 203 (2022).
42. P. A. Vanden Bout, R. L. Dickman and A. L. Plunkett, *The ALMA Telescope: The Story of a Science Mega-Project* 2023.
43. M. Delli Veneri, Ł. Tychoniec, F. Guglielmetti, G. Longo and E. Villard, 3D detection and characterization of ALMA sources through deep learning, *MNRAS* **518**, 3407 (January 2023).
44. F. Guglielmetti, M. Delli Veneri, I. Baronchelli, V. Johnson, P. Arras, C. Blanco, M. Brescia and al., A BRAIN study to tackle imaging with artificial intelligence in the ALMA2030 era (November 2023).
45. The CASA Team, B. Bean and al., Casa, the common astronomy software applications for radio astronomy, *Publications of the Astronomical Society of the Pacific* **134**, p. 114501 (nov 2022).